\documentclass[12pt]{iopart}
\usepackage{times}
\usepackage{mathptmx}
\usepackage{graphicx}
\usepackage[dvips,colorlinks]{hyperref}

\begin{document}

\centerline { \bf RAMANUJAN SUMS FOR SIGNAL PROCESSING OF LOW FREQUENCY NOISE}

\bigskip 

\centerline {M. Planat$^1$, H.C. Rosu$^2$, S. Perrine$^3$} 

\bigskip

\centerline{$^1$ Laboratoire de Physique et M\'etrologie du Oscillateurs du CNRS,}
\centerline{32 Avenue de l'Observatoire, 25044 Besan\c con Cedex, FRANCE}

\centerline{$^2$ Dept. of Appl. Math., IPICyT, Apdo Postal 3-74 Tangamanga, San Luis Potos\'{\i}, MEXICO} 

\centerline{$^3$ 5, Rue du Bon Pasteur, 57070 Metz, FRANCE}

\bigskip

\centerline {Dated: June 2002} 

\bigskip

\bigskip

\centerline{\bf Abstract} 

\noindent
{\scriptsize
An aperiodic (low frequency) spectrum may originate from the error term in the mean value
of an arithmetical function such as M\"obius function or Mangoldt function, which are coding 
sequences for prime numbers. In the discrete Fourier transform the analyzing wave is periodic and not 
well suited to represent the low frequency regime. In place we introduce a new signal processing 
tool based on the Ramanujan sums $c_q(n)$, well adapted to the analysis of arithmetical sequences with many 
resonances $p/q$. The sums are quasi-periodic versus the time $n$ of the resonance and aperiodic versus
the order $q$ of the resonance. New results arise from the use of this Ramanujan-Fourier transform (RFT) in the 
context of arithmetical and experimental signals.}   
 
\bigskip


\bigskip

\section*{0. INTRODUCTION} \vspace{1.5ex}
\lq\lq In this age of computers, it is very natural to replace the
continuous with the finite. One thinks nothing about replacing the
real line $R$ with a finite circle (i.e., a finite ring
${\it Z}/q{\it Z}$) and similarly one replaces the real
Fourier transform with the fast Fourier transform" \cite{Terras}.

In this paper our claim is that the discrete Fourier transform
(and thus the fast Fourier transform or FFT) is well suited to the
analysis of periodic or quasi periodic sequences, but fails to
discover the constructive features of aperiodic sequences, such as
low frequency noise. This claim is not new and led to alternative
time series analysis methods such as Poincar\'{e} maps
\cite{Planat1} (i.e. one dimensional return maps of the form
$x_{n+1}=f(x_n)$ or more general multidimensional maps), fractal
or wavelet analysis methods \cite{Mallat} and autoregressive
moving average (ARMA) models \cite{Shumway} to mention a few.
These methods appeared in diverse contexts: turbulence, financial,
ecological, physiological and astrophysical data. For stochastic
sequences such as $1/f$ electronic noise only small progress was
obtained thanks to these techniques \cite{Planat1}.

Here we introduce still another approach by considering the time
series as an arithmetical sequence, that is a discrete sequence
$x(n),~ n=1\cdots t$, in which generic arithmetical functions
(such as $\sigma(n)$ the sum of divisors of $n$, $\phi(q)$ the
number of irreducible fractions of denominator $q$, the M\"{o}bius
function  $\mu(n)$, or the Mangoldt function $\Lambda(n)$ 
may be hidden.

Recently we published a number of papers which emphasize the
connection between frequency measurements and arithmetic
\cite{Planat1},\cite{Fluc},\cite{APL02}. The standard heterodyne
method, which compares one oscillator of frequency $f(n)$ at time
$n$ to a reference oscillator of frequency $f_0$, leads to
irreducible fraction $p_i/q_i$ of index $i$ given from continued
fraction expansions of $\nu=f(n)/f_0$ and beat signals of
frequencies $F(n)=f_0q_i|\nu-p_i/q_i|$. Jumps between fractions of
index $i,~i\pm1,~i\pm2\cdots$ were clearly identified as a source
of white or $1/f$ frequency noise in such frequency counting
measurements \cite{Fluc}. A phase locked loop was characterized as
well, leading to a possible relationship between $1/f$ noise close
to baseband and arithmetical sequences of prime number theory
\cite{APL02}.

We introduce Ramanujan sums as a new signal processing tool for
these experimental files. In contrast to the discrete Fourier
transform in which the basis functions are all roots of unity
(Sect. \ref{discrete}), the Ramanujan-Fourier transform (RFT) is
defined from powers over the primitive roots of unity (Sect. \ref{Ramanu}). 
We provide a table of known RFT's and emphasize the
newly discovered connection between $1/f$ noise and arithmetic. In
this context considered in Sect. \ref{noise}, a modified Mangoldt
function plays a central role, since it connects the Golden ratio
found in the slope of the FFT to the M\"{o}bius function found in
the structure of the RFT. Concerning the experimental files
considered in Sect. \ref{experim}, galactic nuclei are promising
candidates for the application of the new method.

\section{THE DISCRETE FOURIER TRANSFORM }
\label{discrete} \vspace{1.5ex} The discrete Fourier transform
(DFT) or its fast analogue (the FFT) is a well known signal
processing tool. It extends the conventional Fourier analysis to
sequences with finite period $q$ (for the FFT one takes $q=2^l$,
with $l$ integer).

In the DFT one starts with the roots of unity of the form
$\exp(2i\pi\frac{p}{q})$, $p=1\ldots q$ and the signal analysis is
performed thanks to the $n^{\rm{th}}$ power
\begin{equation}
e_p(n)=\exp(2i\pi\frac{p}{q}n).
\end{equation}
(In the mathematical language one says that $e_p(n)$ is a
character of $G={\it Z}/q{\it Z}$, it is a group
homomorphism from the additive group $G$ into the multiplicative
group of complex numbers of norm $1$). The DFT of the time series
$x(n)$ is defined as
\begin{equation}
\hat{x}(p)=\sum_{n=1}^q x(n)e_p(-n),
\end{equation}
and they are a number of relations such as the inversion formula
\begin{equation}
x(n)=\frac{1}{q}\sum_{p=1}^q\hat{x}(p)e_p(n),
\end{equation}
the Parseval formula (conservation of energy)
\begin{equation}
\sum_{n=1}^q |x(n)|^2=\frac{1}{q}\sum_{p=1}^q|\hat{x}(p)|^2,
\end{equation}
the orthogonality relations between the characters
\begin{eqnarray}
\sum_{n=1}^q e_p(n)e_r(n)=q\delta_p(r)
=\left\{\begin{array}{ll} & q~ \mbox{if}~p\equiv r(mod~q)\\
  & 0~ \mbox{otherwise},\\
\end{array}\right.
\label{orthog2}
\end{eqnarray}
and the convolution formula
\begin{equation}
\widehat{x*y}=\hat{x}\hat{y},
\end{equation}
where $*$ means the convolution.
Generic discrete Fourier transforms are given in table 1.
%
\begin{table}[htbp]
\caption{\label{TableFour}} \centering
\begin{tabular}{|p{2.5cm}|r|}
\hline
$x(n)$&$\hat{x}(p)$\\
\hline
$1$&$q~\delta_0(p)$\\
\hline
$e_l(n)$&$q~\delta_l(p)$\\
\hline
$\delta_l(n)$ & $e_l(-p)$\\
\hline
$\frac{1}{2}$$(\delta _1$+$\delta _{-1})(n)$ & $\cos(2\pi p/q)$\\
\hline 
$L_q(-n)$&$L_q(n)~\hat{L}_q(-1)$\\
\hline
\end{tabular}
\end{table}
In particular, as  it is well known, an oscillating signal
$e_l(n)=\exp(2i\pi \frac{l}{q}n)$ of frequency $l/q$ transforms to
a line at $p=l$ in the DFT spectrum. Inversely a line at $n=l$ in
the time series transforms to an oscillating signal $\exp(-2i\pi
\frac{l}{q}p)$ of frequency $l/q$.

A Gaussian transforms to a Gaussian through the Fourier integral.
Not so well known is that the role of the Gaussian is played by
the Legendre symbol in the context of the DFT
\cite{Terras},\cite{Schroeder}. Let us define the Legendre symbol
$L_q(n)=\left( \frac{n}{q} \right)$ for an odd prime $q$ as
follows
\begin{eqnarray}
\left( \frac{n}{q} \right) =\left\{\begin{array}{lll}
&~0~\mbox{if}~q~\mbox{divides}~n,\\
& +1~\mbox{if}~n~\mbox{is~a~square~modulo~q} \\
&~~~(x^2\equiv n~(mod~q)~\mbox{has~a~solution}),\\
&-1~\mbox{otherwise}.
\end{array}\right.\nonumber\\
\label{equa4}
\end{eqnarray}
There are a number of relations attached to the Legendre symbol
\begin{eqnarray}
&\left( \frac{n}{q} \right) =n^{\frac{q-1}{2}}(mod~q)\nonumber, \\
&\left( \frac{-1}{q} \right)=(-1)^{\frac{q-1}{2}}\nonumber, \\
&\left( \frac{q}{p} \right)\left( \frac{p}{q}
\right)=(-1)^{\frac{p-1}{2}.\frac{q-1}{2}}~\mbox{for~distinct~odd~primes}~p,q\nonumber,\\
&\left( \frac{2}{q} \right) =(-1)^{\frac{q^2-1}{8}}.\nonumber
\end{eqnarray}
The invariant relation for the DFT on ${\it Z}/q{\it Z}$
is
\begin{equation}
\hat{L}_q(-n)=g~L_q(n)~\mbox{with}~g=\hat{L}_q(-1).
\end{equation}
The Fourier coefficient at position $n$ equals the coefficient of
the original sequence up to a constant factor
$g=\hat{L}_q(-1)=\sum_{p=1}^q \left( \frac{p}{q} \right)
\exp(2i\pi \frac{p}{q} )$ and $g^2=(-1)^{\frac{q-1}{2}}q$.
\section{THE RAMANUJAN FOURIER TRANSFORM}
\label{Ramanu} \vspace{1.5ex}
 Ramanujan sums $c_q(n)$ are defined
as the sums of the $n^{\rm{th}}$ powers of the $q^{\rm{th}}$
primitive roots of the unity \cite{Gadiyar},\cite{Young}
\begin{equation}
c_q(n)=\sum_{\begin{array}{cc} \scriptstyle p=1\\ \scriptstyle
(p,q)=1 \end{array}}^q \exp(2i\pi \frac{p}{q} n),
\end{equation}
where $(p,q)=1$ means that $p$ and $q$ are coprimes. It may be
observed that the $c_q(n)$ are the sums over the primitive
characters $e_p(n)$. The sums were introduced by Ramanujan to play
the role of basis functions over which typical arithmetical
functions $x(n)$ may be projected
\begin{equation}
x(n)=\sum_{q=1}^{\infty} x_q c_q(n).
\end{equation}
It should be observed that the infinite expansion with $q
\rightarrow \infty$ reminds the Fourier series analysis, rather
than the discrete Fourier transform that is taken with a finite
$q$. As a typical example the function $\sigma(n)$ (the sum of
divisors of $n$) expands with a RFT coefficient
$\sigma_q=\frac{\pi^2n}{6}\frac{1}{q^2}$, that is
\begin{equation}
\sigma(n)=\frac{\pi^2 n}{6} \{1+\frac{(-1)^n}{2^2}
+\frac{2\cos(2n\pi/3)}{3^2}+\frac{2\cos(n\pi/2)}{4^2}+\cdots\}.
\end{equation}
For functions $x(n)$ having a mean value
\begin{equation}
A_v(x)=\lim_{t \rightarrow \infty} \frac{1}{t}\sum_{n=1}^t x(n),
\end{equation}
one obtains the inversion formula
\begin{equation}
x_q=\frac{1}{\phi(q)}A_v(x(n)c_q(n)). \label{RFT0}
\end{equation}
More general formula have also been derived \cite{Lucht}. In the
rest of the paper the coefficient $x_q$ given in (\ref{RFT0}) will
be we called the Ramanujan-Fourier transform or RFT. It follows
from a number of important relations. There is the multiplicative
property of Ramanujan sums
\begin{equation}
c_{qq'}(n)=c_q(n)c{_{q'}}(n)~\mbox{if}~(q,q')=1,
\end{equation}
and the orthogonality property
\begin{eqnarray}
&\sum_{n=1}^{qq'}c_q(n)c_{q'}(n)=1~\mbox{if}~q\neq q'\nonumber \\
&\sum_{n=1}^q c_q ^2(n)=q \phi(q) ~\mbox{otherwise},
\end{eqnarray}
which reminds us (\ref{orthog2}). It is relatively easy to evaluate
Ramanujan sums from basic functions of number theory. Let us
denote $(q,n)$ the greatest common divisor of $q$ and $n$. Using
the unique prime number decomposition of $q$ and $n$
\begin{eqnarray}
q=\prod_i q_i^{\alpha_i}~~(q_i~\mbox{prime})\nonumber, \\
n=\prod_k n_k^{\beta_k}~~(n_k~\mbox{prime}),\\
\end{eqnarray}
one gets the number $\phi(q)$ of irreducible fractions of
denominator $q$, also called Euler totient function
\begin{equation}
\phi(q)=q\prod_i (1-\frac{1}{q_i}),
\end{equation}
and a coding of prime numbers from the M\"{o}bius function
$\mu(n)$ which is defined as
\begin{eqnarray}
\mu(n)= \left\{\begin{array}{ll}
&0~\mbox{if}~n~\mbox{contains~a~square}~\beta_k>1,\\
&1~\mbox{if}~n=1,\\
&(-1)^k~\mbox{if}~n~\mbox{is~the~product}\\
&~~~~~~~~~\mbox{of}~k~\mbox{distinct~primes}.
\end{array}\right.
\end{eqnarray}
Ramanujan sums are evaluated from\cite{Hardy}
\begin{equation}
c_q(n)=\mu\left( \frac{q}{(q,n)}\right)\frac{\phi(q)}{\phi\left(
\frac{q}{(q,n)}\right)}.
\end{equation}
Note that for $(q,n)=1$, $c_q(n)=\mu(q)$. The first values are
given from
\begin{eqnarray}
&c_1=\overline{1};~ c_2=\overline{-1,1};~
c_3=\overline{-1,-1,2}\nonumber \\
&c_4=\overline{0,-2,0,2}; ~c_5=\overline{-1,-1,-1,4}\cdots
\end{eqnarray}
where the bar indicates the period. For instance $c_3(1)=-1,~
c_3(2)=-1,~c_3(3)=2,~c_3(4)=-1\ldots$ Some generic
Ramanujan-Fourier transforms are given in table 2.
%
\begin{table}[htbp]
\centering
 \caption{\label{TableRam_}}
\begin{tabular}{|p{3cm}|r|}
\hline
$x(n)$&$x_q$\\
\hline
$\frac{\sigma(n)}{n}$&$\frac{\pi^2}{6}\frac{1}{q^2}$\\
\hline
$\frac{\phi(n)}{n}$&$\frac{6}{\pi^2}\frac{\mu(q)}{\phi_2(q)}$\\
\hline
$b(n)=\frac{\phi(n)\Lambda(n)}{n}$&$\frac{\mu(q)}{\phi(q)}$\\
\hline
$C(n)$ & $\left( \frac{\mu(q)}{\phi(q)}\right)^2$\\
\hline
\end{tabular}
\end{table}
In the table the function $\phi_2(q)$ generalizes Euler function
\begin{equation}
\phi_2(q)=q^2\prod_i (1-\frac{1}{q_i^2})~.
\end{equation}
In $b(n)$ the Mangoldt function $\Lambda(n)$ is defined as
\begin{eqnarray}
\Lambda(n)= \left\{\begin{array}{ll} &\ln p~~~\mbox{if}
~n=p^{\alpha}, p~ \mbox{a~prime}\\
&0~~~~~~\mbox{otherwise}.
\end{array}\right.
\end{eqnarray}
According to Hardy and Littlewood (1922) the number of pair of
primes of the form $p, p+h$ is
\begin{equation}
\pi_h(x)\simeq C(h)\frac{x}{\ln^2(x)},
\end{equation}
with
\begin{eqnarray}
C(h)= \left\{\begin{array}{ll}
&2C_2\prod_{p|h}\frac{p-1}{p-2},~~~\mbox{if}~h~\mbox{odd}\\
&0~,~~~~~~~~~~~~~~~~~~\mbox{if}~h~\mbox{even}.
\end{array}\right.
\end{eqnarray}
where $p>2$ is a prime, and the notation $p|h$ means $p$ divides
$h$. The parameter $C_2\simeq 0.660...$ is the twin prime
constant. It was recently conjectured \cite{Gadiyar} that this
problem of prime pairs is also related to an autocorrelation
function from the Wiener-Khintchine formula
\begin{equation}
A_v(b(n)b(n+h))=C(h).
\end{equation}
\begin{figure}[htb]
\centerline{
\includegraphics[width=4.5cm,angle=-90]{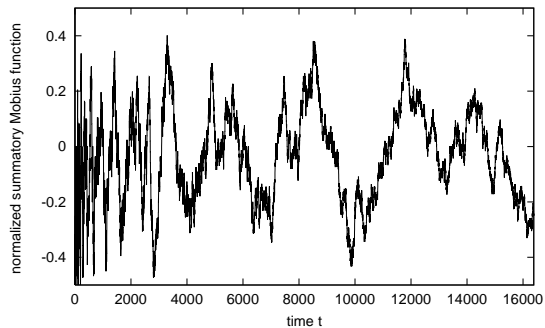}}
\caption{The normalized summatory M\"{o}bius function
$M(t)/t^{1/2}$.} \label{summobius}
\end{figure}
%
\begin{figure}[htb]
\centerline{
\includegraphics[width=4.5cm,angle=-90]{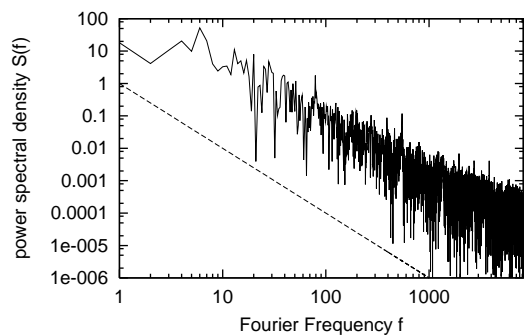}}
\caption{The power spectral density (FFT) of the normalized
summatory M\"{o}bius function $M(t)/t^{1/2}$ in comparison to the
power law $1/f^2$ (dotted line). } \label{FFTmobius}
\end{figure}
%
\begin{figure}[htb]
\centerline{
\includegraphics[width=4.5cm,angle=-90]{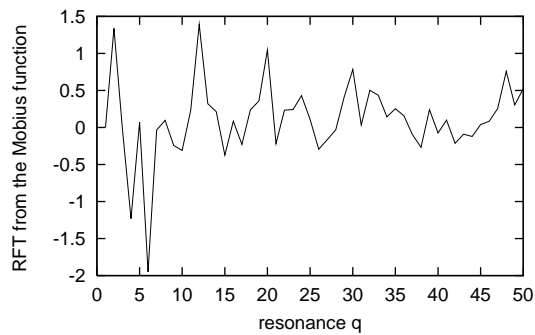}}
\caption{The Ramanujan-Fourier transform (RFT) of the normalized
summatory M\"{o}bius function shown in Fig.1.} \label{RFTmobius}
\end{figure}
%
\section{LOW FREQUENCY NOISE FROM ARITHMETICAL FUNCTIONS}
\label{noise} \vspace{1.5ex} The idea which subtends our new
signal processing is that experimental signals may hide
arithmetical features. It is thus very important to master the low
frequency effects due to generic arithmetical functions such as
M\"{o}bius function, Mangoldt function and so on
\cite{Fluc},\cite{APL02}.

\subsection{On the summatory M\"{o}bius function}
Let us consider the summatory function
\begin{equation}
M(t)=\sum_{n=1}^t
\mu(n)=O(t^{\frac{1}{2}+\epsilon}),~\mbox{whatever}~\epsilon.
\end{equation}
The asymptotic dependance assumes the Riemann hypothesis
\cite{Fluc}. The normalized summatory function $M(t)/t^{1/2}$ is
shown in Fig. \ref{summobius}. The corresponding power spectral
density is in Fig. \ref{FFTmobius}; it looks like the FFT of a
random walk since the slope is close to $-2$.

The RFT of $M(t)/t^{1/2}$ is shown in Fig. \ref{RFTmobius}. There
is no known formula for it, but it shows a signature with well
defined peaks which is reminiscent of the function
$\mu(q)/\phi(q)$ shown below in Fig. \ref{RFTsumnewMangoldt}.

\subsection{Results related to the Mangoldt function}
\begin{figure}[htb]
\centerline{
\includegraphics[width=4.5cm,angle=-90]{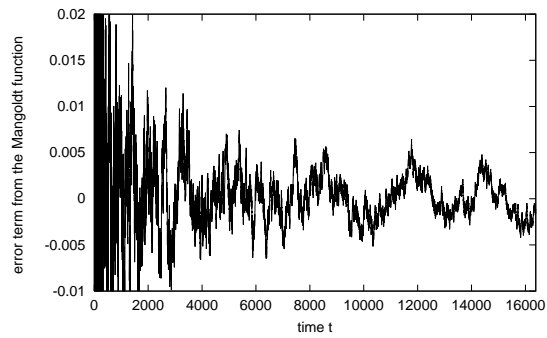}}
\caption{Error term in the Mangoldt function $\Lambda(n)$.}
\label{sumMangoldt}
\end{figure}

\begin{figure}[htb]
\centerline{
\includegraphics[width=5cm,angle=-90]{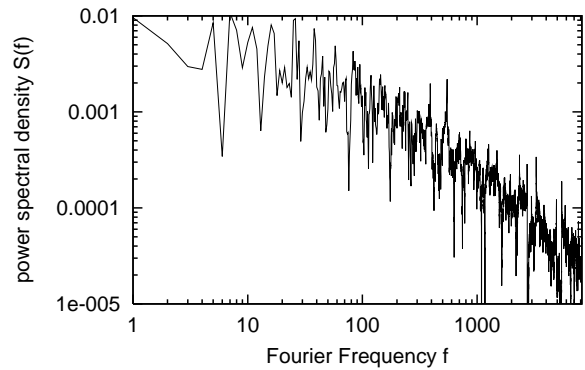}}
\caption{Power spectral density (FFT) of the error term of
Mangoldt function $\Lambda(n)$. } \label{FFTMangoldt}
\end{figure}

Riemann hypothesis can also be studied thanks to the summatory
Mangoldt function
\begin{equation}
\psi(t)=\sum_{n=1}^t \Lambda(t)=t(1+\epsilon_{\psi}(t)).
\end{equation}
The error term represented in Fig. \ref{sumMangoldt} can be
expressed analytically from the singularities (the pole and the
zeros) of the Riemann zeta function \cite{Fluc}. Fig.
\ref{FFTMangoldt} shows the FFT of the error term
$\epsilon_{\psi}(t)$: it roughly behaves as $1/f$ noise.

\begin{figure}[htb]
\centerline{
\includegraphics[width=4.5cm,angle=-90]{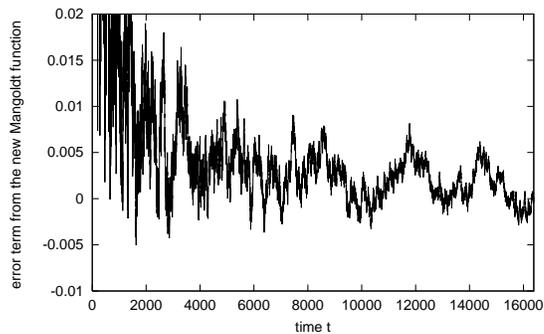}}
\caption{Error term in the new Mangoldt function $b(n)$.}
\label{sumnewMangoldt}
\end{figure}

\begin{figure}[htb]
\centerline{
\includegraphics[width=4.5cm,angle=-90]{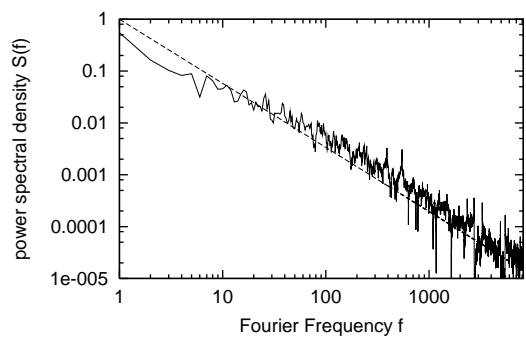}}
\caption{Power spectral density (FFT) of the error term in new
Mangoldt function $b(n)$ in comparison to the power law $1/f^{2
\alpha}$, with $\alpha=(\sqrt{5}-1)/2$, the golden mean. }
\label{FFTsumnewMangoldt}
\end{figure}

Hardy found that the RFT of the modified Mangoldt function
$b(n)=\Lambda(n)\phi(n)/n$ equals $\mu(q)/\phi(q)$. It is thus
interesting to look at the summatory function
\begin{equation}
B(t)=\sum_{n=1}^t \Lambda(n)\phi(n)/n=t(1+\epsilon_B(t)).
\label{mMang}
\end{equation}
The error term in Fig. \ref{sumnewMangoldt} is found to follow
approximately the power law
\begin{equation}
S_B(t) \sim  f^{-2 \alpha}
\end{equation}
with $\alpha=(\sqrt{5}-1)/2=1/(1+1/(1+1/[1+...)))$, as shown in
Fig. \ref{FFTsumnewMangoldt}. This spectrum shows a possible
connection between $\alpha$ and $\mu(q)$ and thus a possible
relationship between the theory of diophantine approximations for
quadratic irrational numbers such as $\alpha$ and prime number
theory. The RFT of $\epsilon_B(t)$ looks similar to the one
$\mu(q)/\phi(q)$ of the new Mangoldt function $b(n)$.

\begin{figure}[htb]
\centerline{
\includegraphics[width=4.5cm,angle=-90]{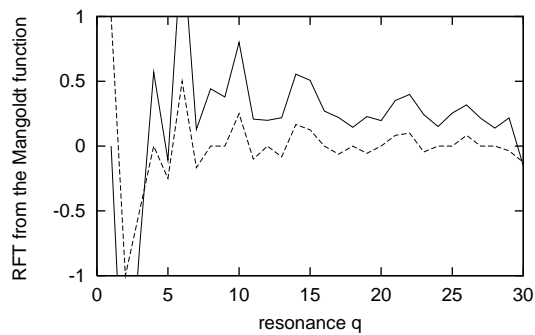}}
\caption{Ramanujan-Fourier transform (RFT) of the error term
(upper curve) of new Mangoldt function $b(n)$ in comparison to the
function $\mu(q)/\phi(q)$(lower curve). }
\label{RFTsumnewMangoldt}
\end{figure}
%
\section{LOW FREQUENCY NOISE FROM EXPERIMENTAL DATA}
\label{experim} \vspace{1.5ex}

Our final goal in using the Ramanujan-Fourier transform is to
discover known arithmetical rules behind experimental sequences.

\subsection{Low frequency noise from galactic nuclei}
\label{AGN}
 We give an example taken from astronomy. The
observation of variability in astronomical systems may lead to
valuable information on the physical nature of the observed
system. In particular Seyfert galaxies are a subset of galaxies
which exhibit evidence for highly energetic phenomena in their
nuclei: they are called active galactic nuclei or AGN. They are
thought to be powered by accretion onto massive black holes at
their centers. X-rays are created mainly in high temperature, high
density regimes, and since matter is fairly transparent to high
energy X-rays, monitoring X-ray emission from AGNs provides a view
into the core and may be used to understand the accretion process
there.

\begin{figure}[htb]
\centerline{
\includegraphics[width=4.5cm,angle=-90]{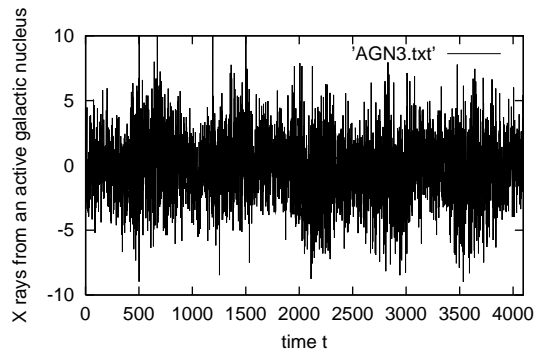}}
\caption{X ray variability from an active galactic nucleus (AGN).}
\label{AGN3}
\end{figure}

\begin{figure}[htb]
\centerline{
\includegraphics[width=4.5cm,angle=-90]{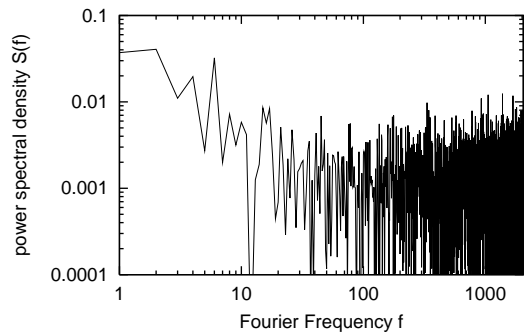}}
\caption{Power spectral density (FFT) of X ray variability from an
AGN. } \label{FFTAGN3}
\end{figure}

\begin{figure}[htb]
\centerline{
\includegraphics[width=4.5cm,angle=-90]{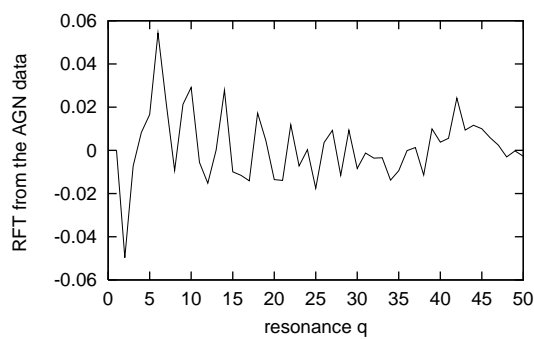}}
\caption{Ramanujan-Fourier transform (RFT) of X ray variability of
an AGN. } \label{RFTAGN3}
\end{figure}

Here we used a sample of data taken from the EXOSAT archive by M.
Koenig and available at http://astro.uni-tuebingen.de/groups/time/
(sample mkn766-85.dat.outZRM) (see Fig. \ref{AGN3}). The power
spectral density exhibits a 1/f low frequency noise as well as
white noise as shown in Fig. \ref{FFTAGN3}. The corresponding RFT
analysis shown in Fig. \ref{RFTAGN3} shows a well defined
signature reminiscent of the RFT signature of Mangoldt function,
that is $\mu(q)/\phi(q)$. That may be an indication that many
resonance processes occur between the black hole and the matter to
be accreted, a process which may be described from prime number
theory.

%
\subsection{Low frequency noise close to phase locking}
\label{PLL}
\begin{figure}[htb]
\centerline{
\includegraphics[width=4cm,angle=-90]{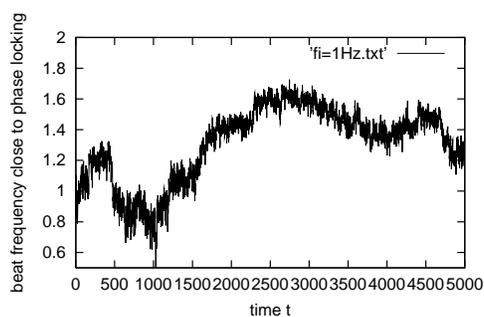}}
\caption{Beat frequency between two radio frequency oscillators
close to phase locking. } \label{fi=1Hz}
\end{figure}

\begin{figure}[htb]
\centerline{
\includegraphics[width=4cm,angle=-90]{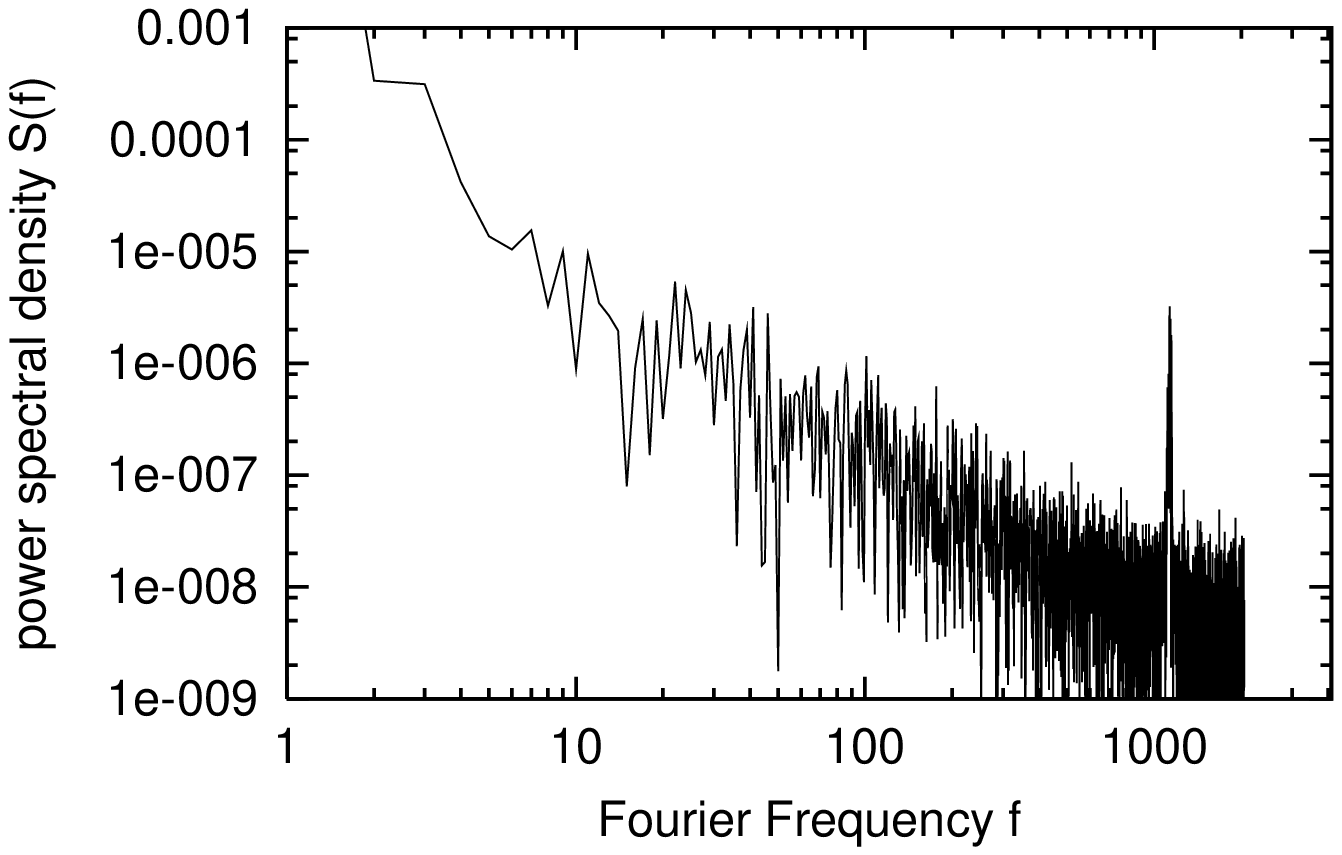}}
\caption{FFT of the beat frequency for two oscillators close to
phase locking. } \label{FFTfi=1Hz}
\end{figure}

\begin{figure}[htb]
\centerline{
\includegraphics[width=4cm,angle=-90]{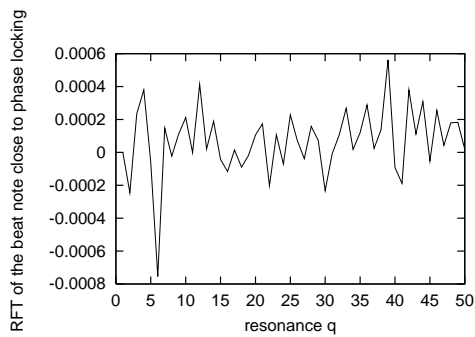}}
\caption{RFT of the beat frequency for two oscillators close to
phase locking. } \label{RFTfi=1Hz}
\end{figure}

 Our second example is taken from the study of
radio frequency oscillators close to phase locking. We recently
demonstrated a relation between phase locking, $1/f$ frequency
noise and prime numbers \cite{APL02}. According to that approach
the coupling coefficient between the oscillators could be
described from a Mangoldt function, leading to de-synchronization
effects and $1/f$ frequency noise. The RFT should be able to
support that conjecture. Figs. \ref{fi=1Hz}, \ref{FFTfi=1Hz} and
\ref{RFTfi=1Hz} show the beat note close to phase locking of $5$
MHz oscillators, the $1/f$ noise calculated from the FFT and the
corresponding RFT.

%
\section{DISCUSSION} There are at least two great challenges in the
study of Ramanujan sums. One can be interested in the extraction
of arithmetical features from experimental files, with the aim to
develop a relevant theory of their randomness. We have in mind the
signals exhibiting $1/f$ noise, since this type of noise still
carries much mystery, in electronics as well as in other fields,
from physics to biology and society.

The RFT signature of $1/f$ noise in a phase locked loop studied in
Sect. \ref{PLL} still doesn't keep one's promise, since we were
unable to relate it to a known arithmetical function. Further work
is required. In contrast, high energy astrophysics seems to be a
relevant field for Ramanujan sums based signal processing. They
may help to derive plausible theories of the strong variability
observed close to Seyfert or other massive galaxies. See the
reference \cite{Rosu} for another application of arithmetic to the
black-hole remote sensing problem.

The other challenge behind Ramanujan sums relates to prime number
theory. We just focused our interest to the relation between $1/f$
noise in communication circuits and the still unproved Riemann
hypothesis\cite{NOAR}. The mean value of the modified Mangoldt
function $b(n)$, introduced in (\ref{mMang}), links Riemann zeros
to the $1/f^{2\alpha}$ noise and to the M\"{o}bius function. This
should follow from generic properties of the modular group
$SL(2,{\it Z})$, the group of $2$ by $2$ matrices of
determinant $1$ with integer coefficients\cite{Perrine}, and to
the statistical physics of Farey spin chains\cite{Knauf}. See also
the link to the theory of Cantorian fractal space
time\cite{Castro}.

\section*{ACKNOWLEDGMENTS}

One author (M. Planat) acknowledges N. Ratier for having pointed
out the paper \cite{Gadiyar} and for his help in programming. He
also acknowledges R. Padma for useful comments on that topic.

\end{document}